\documentclass[%
   reprint,
    twocolumn,
 showpacs,
 showkeys,
 preprintnumbers,
 amsmath,amssymb,
 aps,
   pra,
  longbibliography,
 ]{revtex4-1}

\usepackage{amssymb,amsthm,amsmath}

\theoremstyle{definition}
\newtheorem{definition}{Definition}

\newtheorem{conjecture}{Conjecture}

\theoremstyle{remark}

\usepackage{tikz}
\usepackage[breaklinks=true,colorlinks=true,anchorcolor=blue,citecolor=blue,filecolor=blue,menucolor=blue,pagecolor=blue,urlcolor=blue,linkcolor=blue]{hyperref}
\usepackage{graphicx}
\usepackage{url}

\usepackage{xcolor}

\begin{document}

\title{Unscrambling the Quantum Omelette}


\author{Karl Svozil}
\affiliation{Institute for Theoretical Physics, Vienna
    University of Technology, Wiedner Hauptstra\ss e 8-10/136, A-1040
    Vienna, Austria}
\email{svozil@tuwien.ac.at} \homepage[]{http://tph.tuwien.ac.at/~svozil}


\begin{abstract}
Based on recent theorems about quantum value-indefiniteness it is conjectured that many issues of ``Born's quantum mechanics'' can be overcome by supposing that only a single pure state exists; and that the quantum evolution permutes this state.
\end{abstract}

\maketitle

\section{Ingredients}

The following rather ``iconoclastic'' recasting of quantum mechanics applies to the quantum formalism
as outlined by von Neumann \cite{v-neumann-49}.
It will most likely survive this theory
because the definitions, conventions and results presented apply to a reversible (indeed, bijective)
state evolution,
which amounts to permutations of elements in some state space.
The title is taken from a passage of Jaynes \cite{jaynes-90},
presenting the current quantum mechanical formalism as
{\em ``not purely epistemological; it is a peculiar mixture describing
in part realities of Nature, in part incomplete human information about Nature -- all scrambled up
by Heisenberg and Bohr into an omelette that nobody has seen how to unscramble.''}

What might be the ingredients of such a quantum omelette?
First and foremost, we need to keep in mind that we are dealing with {\em intrinsic self-perception:}
no observer has a ``direct, detached, objective, extrinsic'' viewpoint;
all observers are ``embedded'' in the system they observe (``Cartesian prison'') \cite{bos1,toffoli:79,svozil-94}.

Second, all observations are based on  {\em detector clicks}.
Based on these clicks, and through {\em projections and conventions}
of our mind we reconstruct what we consider the physical universe.
Any inductive (re-)construction of a representation of a universe entirely from ``physical signals''
and, in particular, from detector clicks, is a subtle epistemic and physical task \cite{sum-3,wheeler-89}
involving explicit and implicit conventions and assumptions.
As we do not possess any direct access to the
system other than these clicks  we have to be careful in ascribing physical properties and existence
to anything \cite{stace1}.
Indeed, it must be expected that we are deceived
by our preconceptions, implicit conventions, and subjective expectations and projections.
Jaynes called this
the ``Mind Projection Fallacy'' \cite{jaynes-89,jaynes-90}, pointing out that
{\em ``we are all under an ego-driven temptation to project our private
thoughts out onto the real world, by supposing that the creations of one's own imagination are real
properties of Nature, or that one's own ignorance signifies some kind of indecision on the part of
Nature.´´}
I believe that this ``over-interpretation of empirical data,'' in particular, of detector clicks,
is at the heart of many misconceptions
about quantized systems.

Let us, as a starter, mention some quantum examples of the Mind Projection Fallacy.
First, consider the inclinations \cite{born-26-1} yielding claims \cite{zeil-05_nature_ofQuantum}
of absolute, irreducible indeterminism and randomness,
demanding the ``{\it ex nihilo} emergence of single bits (of information).''
In this orthodox line of thought,
the apparent lack of prediction and control is not merely ``means-relative'' \cite{Myrvold2011237} but ``absolutely irreducible.'' In particular,
the possibility of mere epistemic ignorance, originating from the limited capacities of intrinsic observers,
resulting in ``pragmatic'' propositions that are true ``for all practical purposes'' (FAPP) \cite{bell-a} but strictly false,
is denied.

Rigorously speaking,
any believe in (in-)determinism is {\em provably unprovable} because,
by reduction to recursion theoretic unknowables
(e.g., the halting problem or the rule inference problem \cite{go-67,blum75blum,angluin:83,ad-91,li:92}),
randomness as well as determinism turn out to be undecidable.
That is, one may still be ``inclined to believe in (in-)determinism'' \cite{born-26-1},
and this believe might serve as a good, pragmatic working hypothesis for various tasks;
alas, strictly speaking, any such ``evidence'' is no more compelling than, say, the belief in Santa Claus.

An algorithmic proof can be sketched as follows:
For the sake of an argument against provable indeterminism,
suppose Bob presents Alice a black box,
thereby wrongly claiming that the box contains an oracle for indeterminism, or even absolute randomness.
Alice's challenge is to ``verify'' that this is correct.
As it turns out, Alice's verification task is impossible if she is bound by intrinsic algorithmic means,
because every time Alice has made up her mind
that no algorithm from a particular finite set of algorithms is generating the output of the box,
by diagonalization
Bob can construct a ``faker box algorithm'' which yields a different output than Alice's finite set of algorithms;
thereby giving Alice the wrong illusion of randomness.
With finite physical means the limit of
``all (i.e., a countable infinity of) algorithms'' is impossible to attain.
But even for a finite number of algorithms, their output behavior is FAPP impossible to predict, since
the halting time of a program of fixed length is of the order of the Busy Beaver function of that length,
and therefore grows faster than any computable function thereof \cite{chaitin-bb}.

On the other hand, for the sake of an argument against provable determinism, suppose Bob claims that the box behaves deterministically.
In this case, Alice can be deceived as well; because whenever she claims to know such an algorithm,
by diagonalization
Bob can fabricate another ``faker algorithm'' which behaves exactly as Alice's algorithm until she mentions her claim,
and subsequently behaves differently.
In that way, Alice will never be able to prove determinism.

Of course, the obvious ``solution'' would be to allow Alice to ``screw open Bob's box''
and see whether contained in it there is any ``paper-and-pencil Turing type machinery;''
alas this is not allowed in the intrinsic epistemology.

Other fallacies involve so-called {\em ``experimental proofs of the Kochen-Specker (KS) theorem''}
-- because ``how can you measure a [proof by] contradiction?'' \cite{clifton};
as well as {\em ``experimental proofs of contextuality''} -- what is actually measured are
violations of Boole-Bell type inequalities {\em via}
successive measurements of counterfactual, complementary observables that are not co-measurable \cite{cabello:210401}.
Although contextuality might be {\em  sufficient} to render any experimental records (even beyond quantum correlations \cite{svozil-2011-enough}),
these experiments fall short of any strict test of the {\em necessity} of contextuality.

Still another fallacy is the assumption of the {\em physical (co-)existence of counterfactuals}
(Specker's ``Infuturabilien'' referring to scholastic debates); that is, hypothetical observables
that one could have, but did not measure;
instead some different, complementary, observable has been measured. We shall come back to this issue later.
Finally let me mention the fallacy of supposing that there is some {\em space-time theater}
in which events occur; rather than the ``operationalization'' of space-time {\em via} events \cite{svozil-1996-time,Knuth-Bahreyni}.


\section{Ontological single pure state conjecture}

So, in view of these epistemic limitations and pitfalls,
how might we ``unscramble'' the quantum omelette?
In what follows, the KS and related theorems will be used as a guiding principle.
But first, we need to clarify what constitutes a pure quantum state.
\begin{definition}[State]
Informally, we shall assume that  a
{\em  pure  state} is characterized by
the {\em maximal information} encodable into a physical system.
This can, for instance, be realized by a generalized beam splitter configuration \cite{rzbb}
with an array of detectors; of which only one clicks, the others remain silent.
Formally, a pure quantum state can be represented by a {\em two-valued measure}
either (i) on
 an {\em orthonormal basis};
or (ii) on the spectral decomposition of
a {\em  maximal operator}, from which all commuting orthogonal projectors corresponding to (i) can be functionally derived
(they occur in the spectrum);
or (iii) on
a {\em   context, subalgebra} or {\em block};
or (iv) on the constituents of
 a {\em  unitary transformation} ``encoding'' the basis states (i) by, say, arranging the coordinates of the
basis as either rows or columns in a matrix representation, and singling out one of the basis elements to ``be true.''
\end{definition}

The (strong) KS theorem is usually proved by taking a finite subset of interconnected
(the dimension of the vector space must be three or higher for interconnectivity)
contexts
(or any similar encoding thereof, such as maximal observables, orthogonal bases, or unitary operators),
and by demonstrating that no two-valued measure (interpretable as classical truth assignment)
exists on those structures of observables
if non-contextuality is required -- meaning that the measure is independent of the context.
In a classical (non-contextual) sense, ``somewhere'' in these finite constructions any attempt to
overlay a two-valued measure
-- that is, any enumeration of truth assignments regarding the propositions about outcomes of conceivable measurements --
 must break down due to inconsistencies.
This also occurs, at least for some members of an ensemble, in Boole-Bell-type configurations \cite{peres222}.
Other weak forms of the KS theorem allow two-valued measures,
alas they may be too scarce to, for instance, be able to separate all observables; and to allow a
homeomorphic embedding into Boolean algebras.

A formalism defining partial frame functions,
similar to the one developed in Ref. \cite{2012-incomput-proofsCJ,2013-KstLip}
(instead of the ``holistic'' frame function defined everywhere by
Pitowsky's {\em logical indeterminacy principle} \cite{pitowsky:218,hru-pit-2003})
can, in a particular sense,  be considered an ``improved'' version of the KS theorem
which certifies ``breakdown of (non-contextual) value definiteness'' for any observable
$\vert \textsf{\textbf{b}} \rangle  \langle \textsf{\textbf{b}} \vert$
(associated with the vector $\vert \textsf{\textbf{b}} \rangle$;
from now on, the vector and its associated projector will be used synonymously),
if the
quantum is prepared in a particular state such that the observable $\vert \textsf{\textbf{c}}\rangle$,
which must be non-orthogonal and non-collinear to $\vert \textsf{\textbf{b}} \rangle$, occurs with certainty.
More formally, by considering some finite  construction of interconnected contexts
$\Gamma (C_1,C_2,\ldots ,C_i)$, $i<\infty$,  it turns out that
both possible value assignments
$v( \vert \textsf{\textbf{b}} \rangle )= 0$ as well as
$v( \vert \textsf{\textbf{b}} \rangle )= 1$ are inconsistent with the value assignment $v( \vert \textsf{\textbf{c}} \rangle )= 1$
for any non-orthogonal and non-collinear $\vert \textsf{\textbf{b}} \rangle$.
While, for proof technical reasons,
the Abbott-Calude-Conder-Svozil theorem (ACCS) \cite{2012-incomput-proofsCJ} restricted the angles to
$\sqrt{{5/14}} \le \vert \langle \textsf{\textbf{c}} \mid \textsf{\textbf{b}}\rangle \vert \le  {3/\sqrt{14}}$,
these boundaries have been extended in a recent paper by Abbott, Calude, and the author \cite{2013-KstLip}.

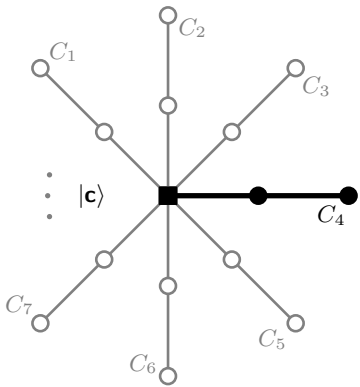
\begin{figure}
\begin{center}
\begin{tikzpicture}  [scale=0.8]
\tikzstyle{every path}=[line width=1pt]
\tikzstyle{every node}=[draw,line width=1pt,inner sep=0]

\tikzstyle{c1}=[rectangle,minimum size=6]

\tikzstyle{d1}=[circle,draw=none,fill,minimum size=2]

\tikzstyle{l7}=[draw=none,circle,minimum size=45]

\draw[gray] (0:0) -- (135:3)
        coordinate[c1,at start] (0)
        coordinate[c1,circle,midway,fill=white] (1)
        coordinate[c1,circle,at end,fill=white,label=35:$C_1$] (2);

\draw[gray] (0.center) -- (90:3)
        coordinate[c1,at start] (0)
        coordinate[c1,circle,midway,fill=white] (3)
        coordinate[c1,circle,at end,fill=white,label=350:$C_2$] (4);

\draw[gray] (0.center) -- (45:3)
        coordinate[c1,at start] (0)
        coordinate[c1,circle,midway,fill=white] (5)
        coordinate[c1,circle,at end,fill=white,label=305:$C_3$] (6);

\draw[gray] (0.center) -- (315:3)
        coordinate[c1,at start] (0)
        coordinate[c1,circle,midway,fill=white] (9)
        coordinate[c1,circle,at end,fill=white,label=215:$C_5$] (10);

\draw[gray] (0.center) -- (270:3)
        coordinate[c1,at start] (0)
        coordinate[c1,circle,midway,fill=white] (11)
        coordinate[c1,circle,at end,fill=white,label=170:$C_6$] (12);

\draw[gray] (0.center) -- (225:3)
        coordinate[c1,at start] (0)
        coordinate[c1,circle,midway,fill=white] (13)
        coordinate[c1,circle,at end,fill=white,label=125:$C_7$] (14);

\draw[black,line width=2pt] (0.center) -- (0:3)
        coordinate[c1,fill,at start] (0)
        coordinate[c1,circle,fill,midway] (7)
        coordinate[c1,circle,fill,at end,label=260:$C_4$] (8);

\coordinate[l7,label=180:$\vert \textsf{\textbf{c}} \rangle$] (0) at (0.center);

\coordinate[d1,gray] (.) at (190:2);
\coordinate[d1,gray] (.) at (180:2);
\coordinate[d1,gray] (.) at (170:2);
\end{tikzpicture}
\end{center}
\caption{(Color online)
Greechie orthogonality diagram of a star-shaped configuration,
representing a common detector observable $\vert \textsf{\textbf{c}}  \rangle \langle  \textsf{\textbf{c}} \vert$ with an overlaid two-valued assignment reflecting $v(\vert \textsf{\textbf{c}}  \rangle)=1$.
It is assumed that the system is prepared in state $C_4$, depicted by a block colored in thick filled black;
all the other (continuity of) contexts are   ``phantom contexts'' colored in gray.
(Compare also Ref. \cite[Fig.~2]{2012-incomput-proofsCJ}.)
}
\label{2012-psiqm-v2}
\end{figure}

In what follows we shall argue that, by explicitly excluding certain  {\em star-shaped configurations of contexts}
characterized by an arbitrary number of orthogonal bases
with one common element (cf. Fig.~\ref{2012-psiqm-v2}),
it is possible to extend  the ACCS theorem to the remaining ``counterfactual observables.''

For the sake of demonstration, consider a configuration of three vectors
$
\vert \textsf{\textbf{a}} \rangle
\perp
\vert \textsf{\textbf{c}} \rangle
\not\perp
\vert \textsf{\textbf{b}} \rangle$,
and a two-valued state
$v(\vert \textsf{\textbf{c}}  \rangle )=1$.
Note that $\vert \textsf{\textbf{a}} \rangle$
lies on the plane (through the origin) orthogonal to $\vert \textsf{\textbf{c}} \rangle$,
whereas
$\vert \textsf{\textbf{b}} \rangle$
lies outside of this orthogonal plane.
In terms of  Greechie orthogonality diagrams \cite{greechie:71},
$
\vert \textsf{\textbf{a}} \rangle
$
as well as
$
\vert \textsf{\textbf{c}} \rangle
$ are contained in a star-shaped configuration of contexts characterized by
the rays perpendicular to some ``true''
$\vert \textsf{\textbf{c}}  \rangle$
with $v( \vert \textsf{\textbf{c}}  \rangle )=1$; whereas
$\vert \textsf{\textbf{b}} \rangle$
lies outside of ``$\vert \textsf{\textbf{c}}  \rangle$'s star.''
For any such observable corresponding to $\vert \textsf{\textbf{b}} \rangle$
there is no consistent non-contextual two-valued state assignment whatsoever.

That is, if $\vert \textsf{\textbf{a}} \rangle$
is orthogonal to $\vert \textsf{\textbf{c}} \rangle$
the value assignment $v(\vert \textsf{\textbf{a}} \rangle)=0$
follows from $v(\vert \textsf{\textbf{c}} \rangle)=1$;
but this latter assignment is inconsistent with either $v(\vert \textsf{\textbf{b}} \rangle)=0$
or $v(\vert \textsf{\textbf{b}} \rangle)=1$ for all $\vert \textsf{\textbf{b}} \rangle$
non-orthogonal and non-collinear to $\vert \textsf{\textbf{c}} \rangle$.
This is also a consequence of Pitowsky's logical indeterminacy principle, which,
given $v(\vert \textsf{\textbf{c}} \rangle)=1$,
does not allow any globally defined two-valued state $v$ which acquires the values
$v(\vert \textsf{\textbf{b}} \rangle)=0$ or
$v(\vert \textsf{\textbf{b}} \rangle)=1$.

For a configuration
$
\vert \textsf{\textbf{a}} \rangle
\not\perp
\vert \textsf{\textbf{c}} \rangle
\not\perp
\vert \textsf{\textbf{b}} \rangle$,
both
$
\vert \textsf{\textbf{a}} \rangle
$
as well as
$
\vert \textsf{\textbf{b}} \rangle
$
lie outside of ``$\vert \textsf{\textbf{c}}  \rangle$'s star,''
and are thus value indefinite.
On the other hand, if we assume $
\vert \textsf{\textbf{a}} \rangle
\perp
\vert \textsf{\textbf{c}} \rangle
\perp
\vert \textsf{\textbf{b}} \rangle$ -- that is, both $\vert \textsf{\textbf{a}} \rangle$ as well as $\vert \textsf{\textbf{b}} \rangle$
are orthogonal to $\vert \textsf{\textbf{c}} \rangle$
(and thus ``in $\vert \textsf{\textbf{c}} \rangle$'s star'') --
$v(\vert \textsf{\textbf{a}} \rangle)=v(\vert \textsf{\textbf{b}} \rangle)=0$, even if they are non-orthogonal.
Hence, given $v(\vert \textsf{\textbf{c}} \rangle)=1$, relative to the KS assumptions,
the only consistent assignments may be made
``inside $\vert \textsf{\textbf{c}} \rangle$'s star.''
``Outside of $\vert \textsf{\textbf{c}} \rangle$'s star''
all ``observables'' are value indefinite (relative to the KS assumptions, including non-contextuality).

How can one utilize these findings?
One immediate possibility is the construction of a {\em quantum random number generator} ``certified by quantum value indefiniteness:''
prepare  $\vert \textsf{\textbf{c}} \rangle$, measure
$\vert \textsf{\textbf{b}} \rangle \langle \textsf{\textbf{b}}\vert$ \cite{2012-incomput-proofsCJ}.

Another intuitive speculation based on the very limited value-definiteness allowed by the KS assumptions
(including non-contextuality)
suggests a foundational principle.
While extensions \cite{2013-KstLip} of the logical indeterminacy principle and the ACCS theorem might never be able to go beyond value indefiniteness
of all but a ``star-shaped'' configuration of contexts depicted in Fig.~\ref{2012-psiqm-v2}, I suggest
to ``get rid'' of even star-shaped configurations
by denying the physical co-existence of
all but one context -- the one in which the quantum has been ``prepared'' -- prior to measurement.
\begin{conjecture}[Ontological single pure state conjecture]
 A quantized system is in a state corresponding to a {\em two-valued measure on a single definite context (orthonormal basis, block, maximal observable, unitary operator). }
In terms of observables, this translates into
{\em ``ontologically there does not exist any observable beyond the observables representing a single definite context.''}
\end{conjecture}

The ontological single pure state conjecture
claims that a single quantum state is a {\em complete}
theoretical representation of a physical system.
Thereby it {\em abandons omni-existence and omniscience:}
it states that all other (even hypothetically and consistently ``value definite'' yet counterfactual) observables
different from the observables associated with the unique state,
and possibly
ascribed to such a system, are not value definite at all.

One should not be ``tricked'' into believing that such value indefinite observables are
``measurable'' just because their alleged ``measurement'' yields outcomes; that is, clicks in detectors that one is inclined to identify with (pre-existing) values.
These {\em outcomes cannot reflect any value definite property of the object prior to measurement}
because, according to the single pure state conjecture,
such a value definite property  simply does not exist.
Rather the detector clicks associated with the ``measurement'' might be a very complex consequence
of {\em ``the complete disposition  of the apparatus''} \cite{bell-66}, as well as of the object, combined.
In contradistinction, orthodox quantum mechanics
treats {\em all potentially conceivable} observables on an {\em equal footing.}

We shall also introduce two other concepts: a {\em phantom context,} and {\em context translation:}
Any context that is not the single context/state (in which the system is prepared) is a {\em phantom context}.
And any  mismatch between the preparation and the measurement may result in the {\em translation}
of the original information encoded in a quantum system into the answer requested,
whereby noise is introduced by the many degrees of freedom of a suitable ``quasi-classical, quasi-chaotic'' measurement apparatus
(for a concrete model, see, for instance, Ref. \cite{Everitt20102809}).

Note that, for this epistemic uncertainty, the resulting stochasticity alone cannot account for greater-than-classical
(sometimes referred to as ``nonlocal'') correlations; rather these reside in the quantum feature of {\em entanglement},
allowing to code information across multiple quanta without defining the (sub-)states of the individual quanta  \cite{zeil-99}.
Thereby, the holistic nature of the quantum entanglement of multipartite system ``creates'' violations
of classical bounds on probabilities and expectations
(see Refs.\cite{toner-bacon-03,svozil-2004-brainteaser} for non-local classical simulations of quantum and even stronger-than-quantum correlations).

For the sake of demonstration of the ontological single pure state conjecture, consider the rule that, under the KS assumptions
(including non-contextuality), for Specker's ``bug'' configuration (Pitowsky's ``cat's cradle'' graph)
of contexts as depicted in Fig.~\ref{2012-psiqm-v2-f2},
if a classical system is prepared in a two-valued state $v(\vert \textsf{\textbf{c}} \rangle )=1$  on the context $C_1$
(i.e. the detector corresponding to observable $\vert \textsf{\textbf{c}} \rangle $ clicks), and with
$v(\vert \textsf{\textbf{a}} \rangle )=v(\vert \textsf{\textbf{d}} \rangle )=0$
(i.e. the detectors corresponding to observables $\vert \textsf{\textbf{a}} \rangle $
and $\vert \textsf{\textbf{d}} \rangle $
do not click),
then
the set of rays
$\Gamma (C_1,C_2,\ldots ,C_7)$ allows only for
$v(\vert \textsf{\textbf{b}} \rangle )=0$; that is,
a detector corresponding to observable $\vert \textsf{\textbf{b}} \rangle $ will not click.
[A rather simple proof by contradiction (wrongly) assumes that $v(\vert \textsf{\textbf{c}} \rangle )=1$
as well as $v(\vert \textsf{\textbf{b}} \rangle )=1$
can coexist consistently, thereby leading to a complete contradiction, since in this case
the value assignment of both link observables for $C_3/C_5$ as well as $C_4/C_5$ have to be  1,
alas these link observables belong to the same block $C_5$.]
That quantum mechanics contradicts this prediction  ``if $v(\vert \textsf{\textbf{c}} \rangle )=1$ then
$v(\vert \textsf{\textbf{b}} \rangle )=0$'' is an immediate consequence of the fact that,
because $\vert \textsf{\textbf{c}} \rangle $ and $\vert \textsf{\textbf{b}} \rangle $ are not in the same block, $\vert \textsf{\textbf{c}} \rangle $ cannot be orthogonal to $\vert \textsf{\textbf{b}} \rangle $,
and hence
$\langle \textsf{\textbf{c}}   \mid  \textsf{\textbf{b}}   \rangle \neq 0$,
implying a non-vanishing probability $\vert \langle \textsf{\textbf{c}}   \mid \textsf{\textbf{b}}   \rangle  \vert^2 \ge 0$.
For a concrete though not unique parametrization of the ``bug'' configuration, see
Fig.~4.2 in Ref.~\cite{svozil-tkadlec}, in which preparation of
$\vert \textsf{\textbf{c}} \rangle \equiv (1/\sqrt{3})\left(\sqrt{2},1,0\right)$ and measurement of
$\vert \textsf{\textbf{b}} \rangle \equiv (1/\sqrt{3})\left(\sqrt{2},-1,0\right)$ implies
a probability of observing  $\vert \textsf{\textbf{b}} \rangle $, given $\vert \textsf{\textbf{c}} \rangle $
of
$\vert (1/\sqrt{3})\left(\sqrt{2},1,0\right) \cdot (1/\sqrt{3})\left(\sqrt{2},-1,0\right)\vert^2 = 1/9$
(and not zero, as predicted from classical non-contextuality).

\begin{figure}
\begin{center}
\begin{tikzpicture} [scale=0.3]

\tikzstyle{every path}=[line width=1pt]
\tikzstyle{c1}=[rectangle,minimum size=8]


\draw[gray]  (7,4) -- (7,-4);
\node [above left, gray] at (7,1) {$C_4$};

\draw[gray]  (7,-4) -- (0,-8) ;
\node [below left, gray] at (4.5,-6.5) {$C_2$};
\draw[gray,fill=white] (7,-4) circle [radius=0.25];
\draw[gray,fill=white] (3.5,-6) circle [radius=0.25];

\draw[gray]  (-7,-4) -- (-7,4);
\node [above right, gray] at (-7,1) {$C_3$};

\draw[gray]  (-7,4) -- (0,8) ;
\node [above right, gray] at (-4.5,6.5) {$C_6$};

\draw[gray]  (0,8) -- (7,4);
\draw[gray,fill=white] (0,8) circle [radius=0.25];
\node [below, gray] at (0,7.5) {$\vert \textsf{\textbf{b}} \rangle $};
\node [above left, gray] at (4.5,6.5) {$C_7$};

\draw[gray]  (-7,0) -- (7,0);
\node [above, gray] at (0,0) {$C_5$};

\draw [line width=2pt,black]  (0,-8) -- (-7,-4)
        coordinate[c1,circle, at end,fill=black] (0)
        coordinate[c1,circle,midway,fill=black] (1)
        coordinate[c1,rectangle,at start,fill=black] (2);
\node [below right, black] at (-4.5,-6.5) {$C_1$};
\node [above, black] at (0,-7.5) {$\vert \textsf{\textbf{c}} \rangle $};
\node [above, black] at (-3.5,-5.5) {$\vert \textsf{\textbf{a}} \rangle $};
\node [above, black] at (-6,-3.75) {$\vert \textsf{\textbf{d}} \rangle $};

\end{tikzpicture}
\end{center}
\caption{(Color online)
``Bug-type'' \cite{Specker-priv} Greechie orthogonality diagram
with an overlaid two-valued assignment reflecting
``$v(\vert \textsf{\textbf{c}} \rangle )=1$ implies $v(\vert \textsf{\textbf{b}} \rangle )=0$.''
This configuration is part of the original
proof of the KS theorem \cite[$\Gamma_1$]{kochen1}.
For concrete coordinatizations, see, for instance, the original paper by Kochen and Specker, as well as Refs.~\cite{svozil-tkadlec,2012-incomput-proofsCJ}.
It is assumed that the system is prepared in state $C_1$, depicted by a block colored in thick filled black;
all the other six remaining contexts $C_2$--$C_7$ are   ``phantom contexts'' colored in gray.
}
\label{2012-psiqm-v2-f2}
\end{figure}

However, since according to  the  single pure state conjecture
only $C_1$ exists, any argument based on the simultaneous co-existence of the
counterfactual phantom contexts $C_2$--$C_7$, and, in particular,
the assumption of a property associated with the counterfactual observable
$\vert \textsf{\textbf{b}} \rangle \langle \textsf{\textbf{b}} \vert $,
is inadequate for quantized systems.


\section{Persistent issues}

\subsection{Do measurements exist?}

Everett  \cite{everett} and  Wigner \cite{wigner:mb}
observed that,
if a unitary (bijective, one-to-one, reversible, Laplacian-type deterministic)
quantum evolution were universally valid,
then any distinction or cut between the observer and the measurement apparatus on the one side,
and the quantum ``object'' on the other side, is not absolute or ontic,
but epistemic, means-relative, subjective and conventional.

Because, suppose that one has defined a cut or difference between some quantum and a ``quasi-classical'' measurement device,
one could, at least in principle and if the unitary quantum evolution is universally valid,
``draw a larger perimeter.'' This ``enlargement'' could contain the entire previous combination,
{\em including} the quantum, the cut, and the measurement device.
If the quantum laws are universally valid, such a quantized system should also undergo
a unitary quantum evolution.
And thus, if quantum mechanics is universally valid,
and if it is governed by unitary, reversible, one-to-one evolution,
how could irreversibility possibly ``emerge'' from reversibility?
FAPP, due to the limitations of the experimenter's capacities
irreversibility may be means-relative;
alas, strictly speaking, it decays into ``thin air.''

Because suppose (wrongly) a hypothetical many-to-one function $h(x)=h(y)$ for $x\neq y$ exists which would somehow
`emerge' from injective functions.
Any such function would have to originate from the domain of one-to-one functions such that,
for all functions $f$ of this class,  $x\neq y$ implies  $f(x)\neq f(y)$
-- or, equivalently, the contrapositive statement (provable by comparison of truth tables)
$f(x) = f(y)$ implies $x = y$,  a clear contradiction with the assumption.

Indeed, by {\em Caylay's theorem}
the {\em unitary transformations} on some Hilbert space ${\mathfrak H}$
form a particular permutation group consisting of those permutations preserving the inner product.
This is a subgroup of the {\em symmetric group}
of all permutations on ${\mathfrak H}$.
So, strictly speaking, any quantum mechanical state evolution amounts to permuting the state,
and therefore leaves no room for ``measurement.''

\subsection{Quantum jellification}

Alas, as Schr\"odinger pointed out, without measurement, the
quantum physicists should be troubled that, due to the coherent superposition
resulting from the co-existence of classically mutually exclusive alternatives,
their  {\em ``surroundings rapidly turning into a quagmire, a sort of a featureless jelly or plasma,
all contours becoming blurred, we ourselves probably becoming jelly fish''}  \cite{schroedinger-interpretation}.

The single pure state conjecture and
the context translation principle
would resolve this conundrum by maintaining that there is only one state ``perceived'' from many
epistemic perspectives \cite{DallaChiara-epistemic}; some of them causing noise which FAPP appears irreducible random to intrinsic observers.
In that sense, the measurement conundrum, with all its variants -- Schr\"odinger's cat and jellyfish
metaphors, as well as the Everett-Wigner critique -- can be ``FAPP-resolved by means-relativity.''

\subsection{Analogues in classical statistical mechanics}

Just as Newtonian physics and electromagnetism appear to be reversible,
the quantum measurement conundrum is characterized by the reversibility of
the unitary quantum evolution.
In this respect, the (ir-)reversibility of quantum measurements
bears some resemblance to statistical mechanics: take, for example, {\em Loschmidt's reversibility paradox}
--
that, for large isolated systems with reversible laws of motion, one should never
observe irreversibility, and thus a decrease in entropy;
or {\em Zermelo's recurrence objection}
--
that, as an isolated system will infinitely often approach its initial
state, its entropy will infinitely often approach the initial entropy and thus cannot constantly
increase;
or the challenge posed by the {\em Loschmidt-Maxwell demon} \cite{maxwell-demon2}.
And just as in statistical mechanics, irreversibility appears to be means-relative \cite{Myrvold2011237} and  FAPP,
yet cannot strictly be true.
Also, the ontic determinism exposed here, accompanied by the epistemic uncertainty induced by context translation,
results in the fact that, at least conceptually and on the most fundamental level,
there need not be any probabilistic description.

\subsection{The epistemic or ontic (non-)existence of mixed states}

From a purely formal point of view,
it is impossible to obtain a mixed state from a pure one.
Because again, any unitary operation amounts to a mere basis transformation or permutation,
and this cannot give rise to any increase in stochasticity or ``ignorance.''
Since the generation of ``ontologically mixed states'' from pure ones would require a many-to-one functional mapping,
we conclude that, just as irreversible measurements, genuine ``ontological mixed states'' originating from pure states cannot exist.
Therefore, any ontological mixed state has to be either carried through from previously existing mixed states (if they exist),
or be FAPP perceived as means-relative.
I would like to challenge anyone with doubts to come
up with a concrete experiment that would ``produce'' a mixed state from a pure one by purely quantum mechanical ``unitary'' means.

\section{Summary}

In summary I hold these conjectures to be true:
a quantum state characterized by the maximal information encoded into a physical system
must formally be represented by some orthonormal basis and a two-valued measure thereon,
or anything encoding it, such as a maximal operator.
At any given moment, a quantized system is in a unique, single such state.
All other contexts are phantom contexts, which have no meaning because they are non-operational at best, and in general misleading.
Randomness does not come about {\it ex nihilo} but by {\em context translation},
whereby the many degrees of freedom of the measurement apparatus contribute
to yield means-relative, FAPP random outcomes.
Finally, also mixed states are means-relative and exist FAPP, but not strictly.

\begin{acknowledgments}
This research has been partly supported by FP7-PEOPLE-2010-IRSES-269151-RANPHYS.
This contribution was done in part during a visiting honorary appointment at the University of Auckland, New Zealand, as well as
at the University of Cagliary, Sardinia, Italy.
Discussions during a {\em LARSIM/QuPa workshop on physics and computation} at the {\it Institut Henri Poincar\'e}, Paris, on June 28-29, 2012,
the {\it Biennial IQSA Conference Quantum Structures 2012} in Cagliari, Sardinia, on July 23-27, 2012,
as well as the conference {\em New Directions in the Foundations of Physics 2013}, in Washington, D.C., on May 10-12, 2013,
where  previous versions of this paper have been presented, are gratefully acknowledged.
I also gratefully acknowledge stimulating discussions with and comments by many peers; in particular, Alastair Abbott, Jeffrey Bub, Cristian S. Calude, William Demopoulos, Christopher Fuchs, and Constantine Tsinakis.
\end{acknowledgments}


%

\end{document}